\documentclass[12pt,twoside]{article}
\usepackage{fleqn,espcrc1}

\usepackage{graphicx}
\usepackage[figuresright]{rotating}

\newcommand{\AmS}{{\protect\the\textfont2
  A\kern-.1667em\lower.5ex\hbox{M}\kern-.125emS}}

\title{High $p_T$ hadron spectra in high-energy heavy-ion collisions}

\author{Xin-Nian Wang \address{Nuclear Science Division, MS 70A-3307\\
Lawrence Berkeley National Laboratory\\
Berkeley, CA 94720, USA }
\thanks{This work was supported by the Director, Office of Energy Research,
Division of Nuclear Physics of the Office of High Energy and Nuclear Physics of
the U.S. Department of Energy under Contract Nos. DE-AC03-76SF00098.}}
       
\begin{document}

\maketitle

\begin{abstract}
We calculate the hadron spectra in high-energy $pp$, $pA$ and $AA$ collisions
within a pQCD parton model. Analyses of
experimental data show that the modification of the spectra in $pA$ and $AA$ as
compared to that in $pp$ collisions is consistent with multiple parton
scattering scenario and is dictated by the transition from absorptive soft
interaction at low $p_T$ to incoherent hard parton scattering at high $p_T$.
This analysis not only sheds new light on the limits to the physics
analysis based on thermal fire-ball models but also provides a more
quantitative baseline results on which one can study the effects of parton
energy loss in the hadron spectra at high $p_T$.
\end{abstract}

\section{Introduction}

Large-$p_T$ partons or jets are good probes of the dense matter
formed in ultra-relativistic heavy-ion collisions \cite{WG92}.
The study of parton energy loss can shed light on the properties of the dense
matter in the early stage of heavy-ion collisions.  Large $p_T$
single-inclusive particle spectra in nuclear collisions are 
sensitive \cite{wang98} to parton energy loss. It is also a crucial test
whether there is any thermalization going on in the initial stage of
heavy-ion collisions.

At low $p_T$ the pQCQ parton model becomes invalid and other alternative
approaches like thermal fire-ball models have to be used, from which one can 
extract the freeze-out temperature, collective radial flow 
velocity and chemical potential \cite{fireball}. Apparently, these thermal
fire-ball models cannot be applied to describe hadron spectra at large
$p_T$. Therefore, it is very important to investigate how well a pQCD parton
model can describe hadron spectra in $pp$ collisions and their modification in
$pA$ and $AA$ collisions and where the transition happens between hard and soft
hadron production. In particular, the impact-parameter or $A$ dependence
of the spectra may be unique to distinguish the parton model from other thermal
fire-ball or hydrodynamic models. One can then at least make a quantitative
conclusion about the validity of different models at different $p_T$ range. The
values of temperature and flow velocity extracted from a fire-ball model
analysis, for example,  will have to be looked at with caution and knowledge
of limitations.

\section{Hadron Spectra in $pp$ Collisions}

In a pQCD parton model, the inclusive particle production cross section in $pp$
collisions is given by
\begin{eqnarray}
  \frac{d\sigma^h_{pp}}{dyd^2p_T}&=&K\sum_{abcd}
  \int dx_a dx_b d^2k_{aT} d^2k_{bT} g_p(k_{aT},Q^2) g_p(k_{bT},Q^2) 
  \nonumber \\ & & f_{a/p}(x_a,Q^2)f_{b/p}(x_b,Q^2) 
  \frac{D^0_{h/c}(z_c,Q^2)}{\pi z_c}
  \frac{d\sigma}{d\hat{t}}(ab\rightarrow cd), \label{eq:nch_pp}
\end{eqnarray}
where $D^0_{h/c}(z_c,Q^2)$ is the fragmentation function of parton $c$
into hadron $h$ as parameterized in \cite{bkk} from $e^+e^-$ data,
$z_c$ is the momentum fraction of a parton jet carried by
a produced hadron. We choose the momentum scale as the transverse
momentum of the produced parton jet $Q=p_T/z_c$. We also use a factor
$K\approx 2$ (unless otherwise specified) to account for higher order QCD
corrections to the jet production cross section.

One normally assumes the initial $k_T$ distribution $g_N(k_T)$ to have a
Gaussian form. In this study we relax the Gaussian form
by assuming a variance which depends on $p_T$ leading effectively to a
non-Gaussian distribution,
\begin{equation}
\langle k^2_T\rangle_N(Q)= 1 ({\rm GeV}^2) + 0.2\alpha_s(Q^2) Q^2.
\end{equation}
The parameters are chosen to reproduce the experimental data at around SPS
energies.

\begin{figure}[htb]
\begin{minipage}[t]{78mm}
\includegraphics[width=2.5 in,height=3.0in]{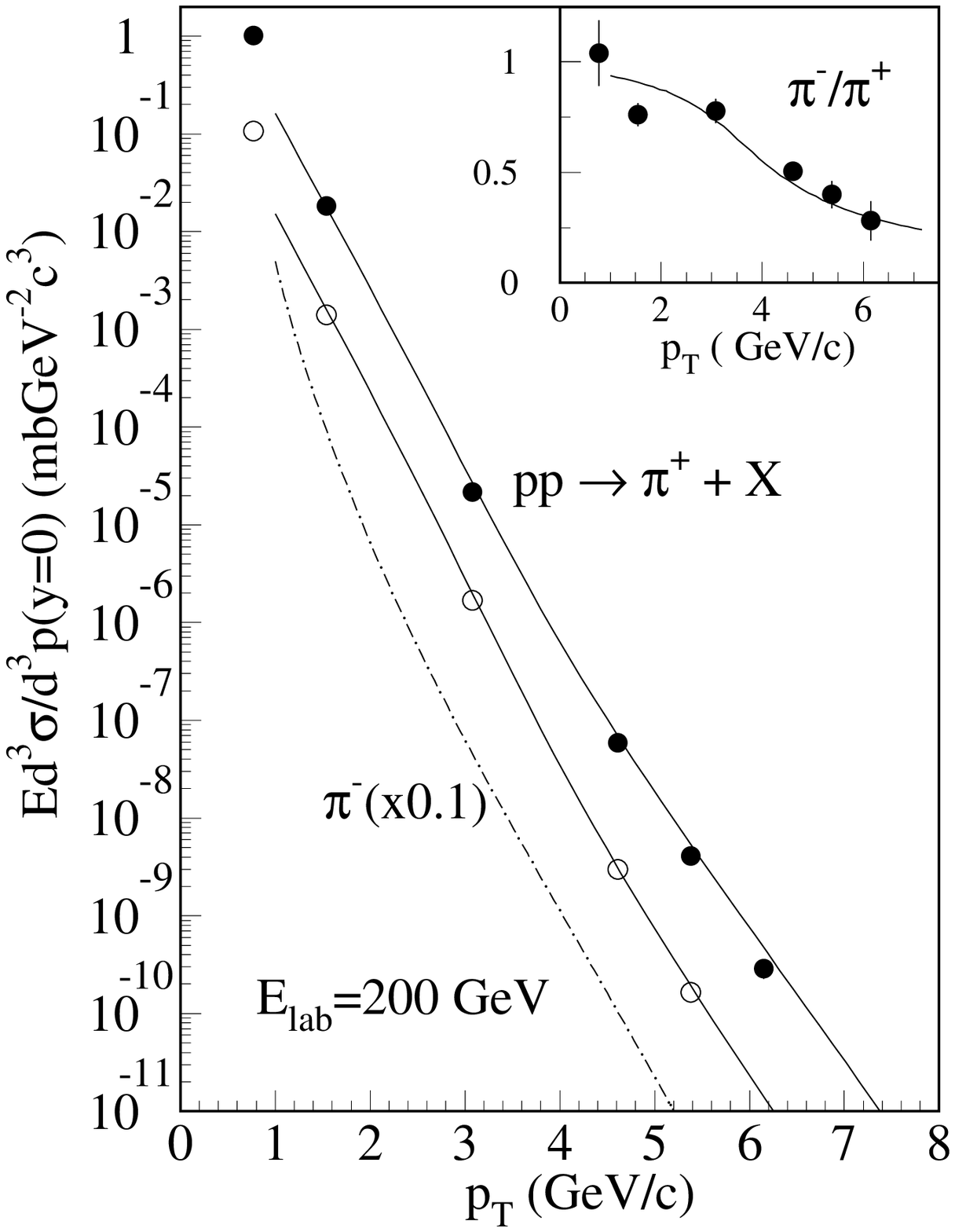}
\caption{ Single-inclusive pion spectra in $p+p$ collisions at 
$\protect E_{\rm lab}=200$ GeV in pQCD parton model with (solid) and without
(dot-dashed) intrinsic $k_T$. Data are from Ref.~\protect\cite{cronin-ex1}.}
\label{figsps1}
\end{minipage}
\hspace{\fill}
\begin{minipage}[t]{78mm}
\includegraphics[width=2.5 in,height=3.0in]{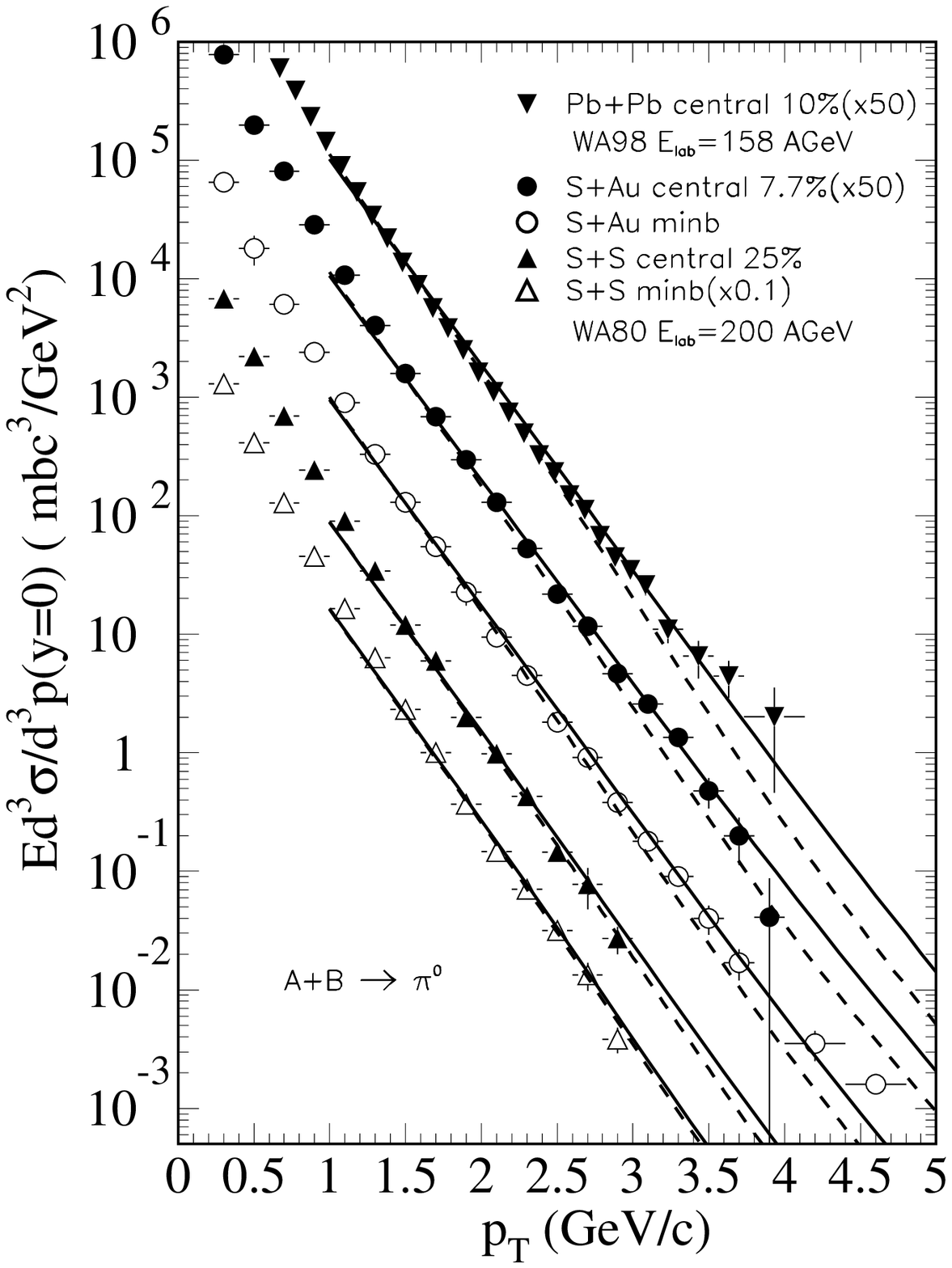}
\caption{ $\pi^0$  spectra in $A+B$
  collisions at the CERN SPS energies in pQCD parton model with (solid) and
  without (dashed) $k_T$ broadening.  Data are from
  Refs.~\protect\cite{wa80,wa98}.} 
\label{figsps11}
\end{minipage}
\end{figure}

    Shown in Fig.~\ref{figsps1} are our calculated
spectra for charged pions as compared to the experimental data
\cite{cronin-ex1} for $p+p$ collisions at $E_{\rm lab}=$200 GeV.
Without the initial $k_T$ smearing the calculations 
significantly underestimate the experimental data. This is because the QCD
spectra are very steep at low energies and even a small amount of
initial $k_T$ could make a big increase to the final spectra. As the
energy increases, the QCD spectra become flatter and small amount of
initial $k_T$ does not change the spectra much. Such parton model calculations
fit very well at energies from $\sqrt{s}=20 - 1800$ GeV for $pp$ and
$p\bar{p}$ collisions \cite{wang99}.
The $p_T$ dependence of $\pi^-/\pi^+$ ratio is another indication of the
dominance of valence quark scattering in large $p_T$ hadron spectra at SPS
energies. 

\section{$pA$ and $AA$ Collisions}

    We assume that the inclusive differential cross
section for large $p_T$ particle production is still given by
hard parton-parton scattering, except that the initial
transverse momentum $k_T$ of the beam partons is broadened. Assuming
that each scattering provide a $k_T$ kick which also has a Gaussian
distribution, we can in effect just change the width of the initial
$k_T$ distribution,
\begin{equation}
\langle k^2_T\rangle_A(Q^2)=\langle k^2_T\rangle_N(Q^2)
    +\delta^2(Q^2)(\nu_A(b) -1).
\end{equation}
The broadening is assumed to be proportional to the number of
scattering $\nu_A(b)$ the projectile suffers inside the nucleus.
We will use the following  $k_T$ broadening per nucleon-nucleon
collision,
\begin{equation}
\delta^2(Q^2)=0.225\frac{\ln^2(Q/{\rm GeV})}{1+\ln(Q/{\rm GeV})}
 \;\;\;{\rm GeV^2}/c^2.
\end{equation}
The $p_T$ dependence of the broadening reflects the fact that
the distribution of soft $k_T$ kick for each scattering does not necessarily
have a Gaussian form. 

The above calculation has been shown to reproduce the nuclear modification of
the hadron spectra in $pA$ data very well \cite{wang99}.
Shown in Fig.~\ref{figsps11} are the calculated inclusive spectra for
produced $\pi_0$ in $A+B$ collisions. The pQCD parton model calculations
with the $k_T$ broadening due to initial multiple scattering (solid lines)
agree with the experimental data (WA80 and WA98) \cite{wa80,wa98} well
very at $p_T$ above 2 GeV/$c$. No parton energy loss has been assumed
in the calculations. The dashed lines are the spectra in
$pp$ collisions at the same energy multiplied by the averaged number of binary
$NN$ collisions given by the nuclear geometrical factor. The difference
between the solid and dashed lines is simply caused by effects
of $k_T$ broadening and nuclear modification of parton distributions
inside nuclei.

\section{$A$ Scaling of Hadron Spectra}

According to the pQCD parton model, the hadron spectra at large $p_T$ should
scale with the number of binary nucleon-nucleon collisions if no nuclear effect
is included. So if one defines a ratio,
\begin{equation}
  R_{AB}(p_T)\equiv \frac{d\sigma^h_{AB}/dyd^2p_T}
  {\langle N_{\rm binary}\rangle d\sigma^h_{pp}/dyd^2p_T} \label{eq:ratio}
\end{equation}
between spectra in $AB$ and $NN$ collisions normalized by the averaged number
of binary collisions $\langle N_{\rm binary}\rangle$, the ratio will be
approximately one for spectra from hard parton collisions.
Because of absorptive processes, low $p_T$ particle production, 
which can be considered as coherent over the dimension of nuclear size, 
has much weaker $A$-dependence.
In the wounded-nucleon model, soft particle production
is proportional to the average number of wounded nucleons, the
above ratio will become
\begin{equation}
R_{AB} \sim 0.5 (1/A^{1/3}+1/B^{1/3})
\end{equation}
So the ratio as defined in Eq.~(\ref{eq:ratio}) will be smaller than one 
at low $p_T$ and larger than one at large $p_T$. Such a general feature 
has been found to be almost universal in both $pA $ and $AB$ collisions.
One interesting feature from this analysis is that the transition between
soft coherent interaction to hard parton scattering happen roughly around
$p_T$=1.5 GeV. This is also the place where hadron spectra in $pp$ collisions
start to deviate from a pure exponential form. 
One can then expect that for spectra above this value of $p_T$
the underlying mechanism of hadron production is dominated by hard
processes.

At even higher $p_T$, the effect of multiple scattering becomes less 
important, so the ratio $R_{AB}$ will approach to 1 again (higher twist
effect should be suppressed by $1/p_T^2$), as shown by the $pA$ data
\cite{wang99}. However, 
at SPS energy, such a feature cannot be be fully revealed because of the 
kinetic limit. One will then only see the initial increase of the ratio
due to the transition from soft to hard processes. Such a change of spectra
from $pp$ to $pA$ and $AA$ collisions in a limited kinetic range looks very
similar to the effect of collective flow in a hydrodynamic model. However,
models motivated by parton scattering have definite $A$-dependence of such a
nuclear modification. Therefore
one should take caution about the values of temperature and flow velocity
extracted from such a fire-ball analysis of the spectra, especially if one
has to reply on the shape of the spectra in the $p_T$ region around 1 GeV.

As one can also observe that the comparison of parton model calculation and the
experiment data does not shown any evidence of parton energy loss
\cite{wang98-2}. 
This implies that the life-time of dense partonic matter could be shorter
than the mean free path of the propagating parton.It also shows that the dense
hadronic matter which has existed for a period of time in the final stage of
heavy-ion collisions does not cause any apparent parton energy loss or jet
quenching. If one observes a dramatic suppression of high $p_T$ hadron spectra
at the BNL RHIC energy as predicted \cite{WG92,wang99,WHS}, then it will
clearly indicate an initial condition very different from what has been reached
at the CERN SPS energy.


\begin{thebibliography}{1}
\bibitem{WG92}X.-N. Wang and M. Gyulassy, Phys. Rev. Lett. 
        {\bf 68}, 1480 (1992).
\bibitem{wang98}X.-N. Wang, Phys. Rev. C {\bf 58}, 2321 (1998).
\bibitem{fireball}U. Heinz, this proceeding.
\bibitem{bkk}J. Binnewies, B. A. Kniehl and G. Kramer, Z. Phys.
        {\bf C}65, 471 (1995).
\bibitem{cronin-ex1}D. Antreasyan, {\it et al.}, Phys. Rev. D{\bf 19}, 764
    (1979).
\bibitem{wang99} X.-N. Wang, nucl-th/9812021
\bibitem{wa80}R. Albrecht {\it et al.}, WA80 Collaboration, Eur. Phys. J. C
        {\bf 5}, 255 (1998). 
\bibitem{wa98}M. M. Aggarwal {\it et al.}, WA98 Collaboration, 
        nucl-ex/9806004 (to be published).
\bibitem{wang98-2}X.-N. Wang, Phys. Rev. Lett. {\bf 81}, 2655 (1998).
\bibitem{WHS}X.-N. Wang, Z. Huang and I. Sarcevic, Phys. Rev. Lett. 
        {\bf 77}, 231 (1996).
\end{thebibliography}
\end{document}